# Polar properties of $Eu_{0.6}Y_{0.4}MnO_3$ ceramics and their magnetic field dependence

J. Agostinho Moreira, A. Almeida, W. S. Ferreira and M. R. Chaves

*IFIMUP and IN- Institute of Nanoscience and Nanotechnology. Departamento de Física da Faculdade de Ciências da Universidade do Porto. Rua do Campo Alegre, 687. 4169-007 Porto. Portugal.*

B. Kundys, R. Ranjith and W. Prellier

*Laboratoire CRISMAT, UMR 6508 CNRS/ENSICAEN, 6 bd du Maréchal Juin
F-14050 CAEN Cedex 4 - France.*

S. M. F. Vilela and P. B. Tavares

*Centro de Química. Universidade de Trás-os-Montes e Alto Douro. Apartado 1013, 5001-801. Vila Real. Portugal.*

*e-mail: jamoreir@fc.up.pt

## Abstract

$Eu_{1-x}Y_xMnO_3$ exhibits, unlike other magnetoelectric systems, very distinctive features. Its magnetoelectric properties is driven by the magnetic spin of the $Mn^{3+}$ ion, but they can be drastically changed by varying the content of $Y^{3+}$, which it does not carry any magnetic moment. Though the x = 0.40 composition has been studied extensively, some basic questions still remain to be thoroughly understood. Thus, this work is aimed at studying some of its polar properties and their magnetic field dependence as well. The experimental results here reported have shown that this material is very easily polarisable under external electric fields, and so, whenever the polarization is obtained from time integration of the displacement currents, an induced polarization is superposed to the spontaneous one, eventually masking the occurrence of ferroelectricity.

We have found clear evidence for the influence of a magnetic field in the polar properties of $Eu_{0.6}Y_{0.4}MnO_3$. The study of electric polarization of $Eu_{0.6}Y_{0.4}MnO_3$ under an external magnetic field yields a value with the same order of magnitude of the remanent polarization determined from polarization reversal experiments. The comparison of the magnetic induced changes on the polarization obtained in polycrystalline samples and single crystals confirms the threshold magnetic field value for the polarization rotation from the a- to the c-direction, and evidencing the importance of the granular nature of the samples in the polar response to magnetic field.

# I. INTRODUCTION

The magnetoelectric effect has attracted a lot of attention motivated by the very interesting fundamental issues that are put forward, as well as by the need of novel memory and sensor devices, where the polarization is controlled by a magnetic field or vice-versa.[1,2] Among the materials presenting a magnetoelectric effect, orthorhombic ReMnO$_3$ (with Re a rare earth ion) have drawn a particular attention in the last years, but some aspects concerning the coupling between electric polarization and magnetic properties are not completely cleared out. In particular, the family of orthorhombic compounds Eu$_{1-x}$Y$_x$MnO$_3$, with x < 0.55 is very interesting, because these materials exhibit a rich variety of phase transitions between incommensurate and commensurate antiferromagnetic phases, some of them with a ferroelectric character, depending on the x value.[3] The fine-tuning of the physical properties through the controlled doping with the non-magnetic Y$^{3+}$ ion, without increasing both magnetic and structural complexity, enables a more direct approach for the understanding of those microscopic mechanisms, which lead to the magnetoelectric effect.

J. Hemberger *et al*,[3] Ivanov *et al*,[4] and Y. Yamasaki *et al*[5] have proposed (*x,T*) phase diagrams, for Eu$_{1-x}$Y$_x$MnO$_3$ single crystals, with 0 ≤ x < 0.55, obtained by using both identical and complementary experimental techniques. Although the phase diagrams proposed by these authors present discrepancies concerning the magnetic phase sequence and the ferroelectric properties for 0.15 < x < 0.25, there is a quite good agreement concerning the phase sequence for 0.25 < x < 0.55. For this range of compositions several main issues have been reported. Ferroelectricity has an improper character, and it likely stems from the inverse effect of the Dzyaloshinskii-Moriya interactions.[5,6,7] A strong magnetoelectric coupling has been evidenced from the effect of applied magnetic fields on their ferroelectric and optical properties.[5,8,9] A synchrotron radiation study has shown that the antiferromagnetic phases have modulated structures,[5] which apparently stem from the competition from both ferro- and antiferromagnetic interactions. Moreover, their electric polarization occurs in a direction perpendicular to its magnetic modulation wave vector.[5,8,9]

Particularly, the $Eu_{0.6}Y_{0.4}MnO_3$ composition has been intensively studied, as it exhibits very interesting and rather unique properties. The phase sequence of $Eu_{0.6}Y_{0.4}MnO_3$, depicted in Figure 1, will be reviewed in the following.

According to the results reported in Refs. 3 and 5, $Eu_{0.6}Y_{0.4}MnO_3$ undergoes a paramagnetic-paraelectric to a collinear sinusoidal modulated antiferromagnetic phase transition at $T_N = 46$ K, hereafter designated as AMF1. The wave vector $q_l$ of the lattice modulation, which is related to the wave vector of the magnetic modulation through $q_l = 2q_m$, is approximately 0.58 down to 10K, presenting a very faint anomaly around 24K.[5] At $T_1 = 24K$, this compound undergoes another phase transition, becoming antiferromagnetic (AFM4), with a presumable cycloidal spin arrangement in the *bc*-plane, with the modulation vector along the *b*-axis.[5] This compound exhibits between $T_1 = 24$ K and $T_2 = 22$ K, a small spontaneous polarization with components along both *a*- and *c*-axes, with values of the order of 100 $\mu C/m^2$.[5,8] At $T_2 = 22$ K, the orientation of the cycloidal plane changes from the *bc* to the *ac*-plane, defining another antiferromagnetic spin arrangement phase (AFM2).[9] In this phase, the component $P_a(T)$ of the polarization considerably increases, attaining a saturation value of about 1200 $\mu C/m^2$, while the $P_c(T)$ component decreases drastically below $T_2$.[5] Noda *et al* [8] found similar results for $Eu_{0.595}Y_{0.405}MnO_3$, but $P_c(T)$ is only different from zero in the interval 21 K – 25 K. These results have been understood as a spontaneous rotation of the electric polarization from *c* to *a* direction, by decreasing the temperature, even at zero magnetic field. This behavior contrasts with the one obtained for $TbMnO_3$, in which an external magnetic field is needed in order to induce a polarization rotation.[10] Noda *et al* [8] and Murakawa *et al* [9] have shown that an applied magnetic field along the *a*-axis of $Eu_{0.55}Y_{0.45}MnO_3$ induces a direction switching of the electric polarization, from the *a*-component to the *c*-component, accompanied by an impressive decrease of the electric polarization. This direction switching has been attributed to the magnetic transition of the Mn 3d basal spins alone.

Actually though, a large effort has been made to understand its physical properties, there is still intriguing fundamental questions regarding the nature and origin of ferroelectricity occurring in both AFM4 and AFM2 phases of $Eu_{0.6}Y_{0.4}MnO_3$. First, it is not clear why in the works referred to above, unusual high electric fields were used to obtain an electric polarization, though it is well know that for measuring a spontaneous polarization much lower electric fields are actually required.[11] Moreover, in Y-doped $EuMnO_3$, $Y^{3+}$ ion distribution can lead to an easily induced electric polarization, even for low polarizing electric fields, which can mask the actual spontaneous polarization, thus turning out the search for ferroelectricity a much more demanding task.[12]

This work is aimed at studying both the nature and origin of the electric polarization occurring in the low temperature antiferromagnetic phase transitions of $Eu_{0.6}Y_{0.4}MnO_3$. After, a detailed characterization of its physical properties had been made, by performing both dielectric and magnetic measurements, the remanent polarization at low temperatures was determined by polarization reversal measurements. The results were subsequently compared with those ones regarding the electric polarization calculated by integrating the displacement currents obtained in heating runs, after cooling the samples under different fixed dc electric fields. This comparison put in evidence an impressive contribution of the induced polarization to the overall polar response. Our study is still focussed on the effect of magnetic fields over the dielectric constant and electric polarization of $Eu_{0.6}Y_{0.4}MnO_3$, addressed in particular to understanding both the magnetodielectric coupling, and the spin-reorientation transition regarding the occurrence of induced electric polarization, inside the antiferromagnetic phases.

Unlike in previous works, the results reported here were obtained in high quality $Eu_{0.6}Y_{0.4}MnO_3$ ceramics instead of single crystals. Despite the granular nature of our samples, the most relevant features observed in $Eu_{0.6}Y_{0.4}MnO_3$ single crystals were very well corroborated, except for both the shape and relative magnitude of the response to external fields. This was especially noticeable, when magnetically induced polarization changes were considered.

## II. EXPERIMENTAL

High quality $Eu_{0.6}Y_{0.4}MnO_3$ ceramics were prepared by the sol-gel combustion method. Details of the sample processing are available in Ref. 13 and 14. The phase purity, the crystallographic and the microstuctural characterization of the ceramic samples were checked using X-ray powder diffraction and scanning electron microscopy with energy dispersive spectroscopy. The Rietveld analysis of the x-ray diffraction data shows the absence of secondary phases, with occupancy factors converging to the nominal composition of the samples. This result was also confirmed by energy dispersion spectroscopy. Scanning electron microscopy analysis reveals in both systems a typical ceramic microstructure with regular shaped crystal grains ranging from 3 up to 10 μm in diameter. A detailed study in $EuMnO_3$ and $GdMnO_3$ ceramics prepared in this way, led to results very similar to the ones obtained in the corresponding single crystals.[15]

Low-field dc magnetization measurements were carried out using a commercial superconducting quantum interference SQUID magnetometer, in the temperature range 4 K – 300 K, with a resolution better than $5 \times 10^{-7}$ emu.

Rectangular shape samples were prepared from ceramic pellets, and gold electrodes were deposited using the evaporation method. The complex dielectric constant was measured with a HP4284A impedance analyzer, in the 10K – 300K temperature range, under an ac electric field of amplitude 100 V/m, in the frequency range 10 kHz to 1 MHz. The displacement currents were measured as a function of temperature, with a standard short-circuit method, using a Keithley electrometer, with 0.5 pA resolution, while keeping a fixed heating temperature rate. The temperature dependence of the corresponding polarisation was obtained by the time integration of the current density. The sample temperature was measured with accuracy better than 0.1 K. The study of polarization reversal, P(E), was done between 45 K and 10 K, using a modified Sawyer-Tower circuit. In order to prevent any dynamical response from masking the actual domain reversal, we have chosen to perform the measurements of P(E) at enough low frequency. As the P(E) relations did not change with frequency below 1 Hz, we took 300 mHz as the operating frequency.

Magnetodielectric and magnetic field dependence of the polarization measurements were carried out in a PPMS Quantum Design cryostat. The capacitance was measured using Agilent 4248A RLC bridge at different frequencies. The polarization was measured with a Keithley 6517A electrometer. For the study of the magnetic field dependence of the electric polarization, the sample was previously cooled down to 10 K, under an electric field of 78 kV/m. Then, the electric field was removed, and the time dependence of the electric polarization was measured for 1000 sec, until obtaining polarization stability. Afterwards, the sample was heated up to 14 K and the time dependence of the electric polarization was recorded, with the simultaneous magnetic field oscillation of 90 kOe amplitude.

## III. EXPERIMENTAL RESULTS AND DISCUSSION

### a. Dielectric and magnetic properties

The temperature dependence of the real ($\varepsilon'$) and imaginary ($\varepsilon''$) parts of the complex dielectric constant, measured at several fixed frequencies, is shown in Figures 2(a) and 2(b), respectively. ε'(T) presents two anomalies, a shoulder at $T_1 = 24$ K that marks the AFM1-AFM4 phase transition, and a peak at $T_2 = 20$ K that marks the AFM4-AFM2 phase transition. ε''(T) shows two peaks at $T_1$ and at $T_2$, and a broad anomaly strongly dependent on the frequency, in the 10 kHz – 1 MHz range, suggesting the existence of a relaxation process. This relaxation process was analysed with a Debye model with a single relaxation time, assuming an Arrhenius law. In the scope of this model, ε''(T) is given by:[16]

$$\varepsilon''(T,\omega) = \frac{\Delta\varepsilon}{2\cosh\left[\frac{U}{k_B}\left(\frac{1}{T} - \frac{1}{T_M}\right)\right]}, \qquad (1)$$

where $\Delta\varepsilon$ is the dielectric strength, U is the activation energy associated with the dielectric relaxation process, $k_B$ is the Boltzmann constant, and $T_M$ is the temperature of the maximum value of ε''(T) for a fixed frequency. In the fitting procedure, we assume that the main temperature dependence in Equation (1) comes from the argument of the hyperbolic function. As it can be seen in Figure 2(a), the values of ε' is nearly temperature independent in the temperature range 40 K – 100 K, so $\Delta\varepsilon$ can be considered as a constant. Figure 3 shows the natural logarithm of the relaxation time τ as a function of the inverse of the temperature. A linear relation between ln(τ) and 1/T between 41 K – 83 K is observed. For T ≈ 40 K, a change of slope is observed. The value of the activation energy, deduced from the best fit of an Arrhenius law to the experimental data is displayed in Figure 2 for 41 < T < 83 K, is 0.020 ± 0.003 eV.

Figure 2(c) shows the temperature dependence of the induced specific magnetization, measured under an applied magnetic field of 50 Oe, in a heating run after cooling the samples under zero field (curve I), or in a heating run after cooling the sample with a dc magnetic field of 50 Oe (curve II). Both curves show a maximum at $T_N = 46$ K, signalizing the transition into the AFM1 phase, in good agreement with previous published results in single crystals as referred to above. Below $T_N$, we observed that the induced magnetization is strongly dependent on how the sample is cooled down. In fact, upon cooling the sample under an external magnetic field of 50 Oe, a significant deviation of both curves is revealed, reaching a 20% difference at 5K. This result evidences some kind of disorder associated with the spin subsystem. This assumption is supported by the incommensurate character of both lattice and magnetic structure in $Eu_{0.6}Y_{0.4}MnO_3$ for $T < T_N$, whose origin has been attributed to ferromagnetic and antiferromagnetic competitive interactions. Curve (II) reaches a saturation just below $T_2$, signaling the onset of the AFM2 phase.

b. Polarization reversal

Figure 4 shows some P(E) relations for $Eu_{0.6}Y_{0.4}MnO_3$, obtained at 330 mHz frequency by applying ac electric fields up to ±1000 kV/m, at different fixed temperatures. For temperatures higher than 30K, a linear relation between P and E confirms the paraelectric character of the AFM1 phase. For temperatures below 25K, we observed hysteresis loops with a very elongated shape. These results, together with the ones obtained for $\varepsilon'(T)$, clearly reveal, in good agreement with the results obtained in single crystals, that $Eu_{0.6}Y_{0.4}MnO_3$ is ferroelectric between 23 K and, at least, 7 K.[4,5] The remanent polarization $P_r(T)$ presents two maxima both at 20 K and 15 K, as can be seen in Figure 5. $Eu_{0.6}Y_{0.4}MnO_3$ is still ferroelectric below 15 K, though its polarization starts to decrease by decreasing temperature until 7 K. It is important to stress that the remanent polarization is much smaller than in common ferroelectrics, reaching a maximum value of just 55 $\mu C/m^2$. From the P(E) relations, it is clearly shown that the electric polarization does never reach saturation, even for electric fields up to 10 kV/m, which evidences that $Eu_{0.6}Y_{0.4}MnO_3$ is a very easily polarisable material.

c. Displacement currents

Figure 6 shows the temperature dependence of the current density, J(T), obtained from sequential thermal cycles, for different values of the polarizing electric field as follows: the current was measured in heating runs with a temperature rate of 2 K/min, after cooling the sample at several fixed dc electric fields. In Figure 6 we can observe two groups of anomalies. Below 30 K a double peak anomaly is clearly associated with the ferroelectric character of the AFM4 and AFM2 magnetic phases, and it will be analysed later on. The anomalies above 30 K have not a ferroelectric character and are associated with a field-induced polarization whose nature can be identified by an appropriate analysis as referred just below. Above 50 K, we can see three maxima rather closed to each other. Just the larger one will be analysed in the following. The nature of the microscopic mechanism underlying the referred anomaly observed in J(T), can be studied by chosen different cooling/heating conditions, as temperature rate and electric field intensity. The amplitude of this peak ($J_M$) is strongly dependent on the magnitude of the applied electric field, and the temperature ($T_M$) where its maximum occurs, remains independent of the magnitude of the electric field, as we can see in Figure 6. Figure 7 shows the temperature dependence of the current density, measured in heating runs for different heating rates, after cooling the sample under a dc electric field of 408 kV/m. Unlike the behaviour of the low temperature anomalies, the temperatures of the maxima, corresponding to the high temperature peaks, increase by increasing the heating rate. The overall behaviour of the anomalies observed above 50 K provides evidence for their dipolar relaxation origin, as follows from the predictions of the Bucci-Fieschi model.[17] In the current literature, these currents are designated by thermally stimulated depolarization currents, and can be expressed as:[17]

$$J(T) = \frac{P_e(T_p)}{\tau_0} \exp\left(\frac{-U}{k_B T}\right) \exp\left[-\frac{1}{q\tau_0} \int_{T_0}^{T} \exp\left(\frac{-U}{k_B T}\right) dt'\right], \qquad (2)$$

where $P_e$ is the equilibrium polarization reached at the polarizing temperature $T_p$, and it depends on the polarizing electric field, $\tau_o$ the relaxation time at infinite temperatures, $U$ the activation energy, $q$ the temperature rate, and $k_B$ the Boltzmann constant. The temperature dependence of the relaxation parameters was found by fitting the straightforward expression for J(T), presented in Ref. 17, to the experimental data.

Figure 8 shows the equilibrium polarization $P_e$ obtained from the best fit of the Equation (2) to the experimental results, concerning the peak under study. As we can see, $P_e$ is proportional to the electric field magnitude. The average value of the activation energy ($U$) associated with the relaxation process is 0.07 ± 0.01 eV (see Figure 8). The relaxation time at infinite temperatures ($\tau_o$) is approximately 0.015 ± 0.002 s.

It is worth to stress that in the temperature range 40 K – 150 K we found relaxation processes, detected through both the study of ε''(T,ω) and displacement currents. Although, as their activation energies are quite different, it is apparent that they should correspond to independent relaxations mechanisms.

The current density displayed in Figure 9 was obtained by cooling the sample from room temperature to 50 K, under a zero electric field, and then the sample was cooled again from 50 K to 7K, under an applied electric field. At 7 K, and after removing the electric field, the electrodes were short-circuited for 30 minutes, and the sample heated from 7 K to 50 K at a rate of 5 K/min. For electric fields lower than 408 kV/m, we observe one sharp anomaly and also a broad anomaly. For electric fields higher than that value, one shoulder at lowest temperatures comes up, whose origin is not known yet. As in ceramics there is a random orientation of their different grains, we observe in the same experiment the whole electric polarization response, which adds up the electric polarizations of both AFM2 and AFM4 phases. The sharp anomaly around 24 K corresponds very clearly to the onset of the AFM4 phase, polar along the *c*-axis, and the broad anomaly correspond to the onset of the AFM2 phase, polar along the *a*-axis. From the shape of the current density curves, metaestability at the merging border of those two phases is expected. Figure 9(b) shows the electric polarization obtained by time integration of J(T), depicted in Figure 9(a). The saturation value of the electric polarization increases as the poling electric field increases, except for E > 918 kV/m. For E = 918 kV/m, the value of the saturation polarization is 380 $\mu C/m^2$.

The inset of Figure 9(b) shows a comparison between the remanent polarization recalled from Figure 5, and the polarisation obtained from the time integration of the displacement current, measured after cooling the sample under a polarizing electric field of 1 kV/m. We have used this field intensity because the induced component does not totally mask the spontaneous polarization. Despite the different magnitude of both results, there is a clear similarity between the overall temperature dependence of the two curves. This feature yields two main conclusions. First, the measurement of the spontaneous polarization obtained from the time integration of displacement currents, has actually to be done in zero field conditions or with a very small polarizing field. Secondly, for higher values of the field a very strong induced component emerges, much larger than the spontaneous polarization, and thus preventing the actual ferroelectric nature of the phase to be firmly settled.

In order to study at what extent the dipolar relaxation above 50K could modify the properties of the ferroelectric phases, we have used a peak cleaning technique as follows:[17] the sample was cooled from 200 K to 40 K under an applied electric field of 408 kV/m; at 40 K the field was removed and the electrodes short-circuited; then, the current was measured in a cooling run from 40 K to 10 K, and afterwards in a heating run from 10 K to 200K, with a temperature rate of 2 K/min. Figure 10(a) shows the experimental results obtained. On cooling, a double peak appears at $T_2$ = 23K indicating the building up of an electric polarization, even in the absence of an applied electric field between 40 K and 10 K. The time integration of this current density gives a polarization of 200 $\mu C/m^2$ at 10 K, well above the value of the remanent polarization obtained from hysteresis loops. So, by cooling the sample, between 200K and 40K under a 408 kV/m electric field, electric dipoles associated with the high temperature relaxations of the sample are oriented, giving rise to an internal electric field, which determines the direction of alignment of both spontaneous and induced dipoles below $T_2$, which have an opposite direction to the one revealed at high temperatures. On a heating run, below 50 K, as expected, the sign of the current density is reversed, and presents only a well defined peak at $T_2$, associated with the transition from the AFM4 to the AFM1 phases. Above 50 K, the results obtained are very similar to those ones displayed in Figure 6, as expected.

Figure 10(b) shows the current density as a function of the temperature, measured in a heating run with a temperature rate of 2 K/min, after cooling the sample with an applied electric field of 408 kV/m, between 40 K and 10 K, and short-circuiting the sample at 10 K during 30 minutes. Above $T_2$ = 24K, as expected, there is no spontaneous polarization, in good agreement with the results obtained from polarization reversal.

d. Polar and dielectric properties under magnetic fields

To check the existence of a magnetodielectric effect in $Eu_{0.6}Y_{0.4}MnO_3$, a measurement of the dielectric constant $\varepsilon'(T)$ was carried out under a magnetic field of 20 kOe, 50 kOe, and 90 kOe, which is parallel to the direction of the measuring ac electric field, at a frequency of 100 kHz. The results are displayed on Figure 11(a). The dielectric peak becomes smaller, and is shifted towards higher temperatures under an external magnetic field. Similar behaviour has been observed in other compounds and is linked to the effect of a magnetic field in the spin configuration, responsible for the electric polarization phenomenon.[18]

The relative value of the magnetodielectric effect determined as $\varepsilon'(B=0T)$- $\varepsilon'(B=9T)$, which is generally proportional to magnetoelectric susceptibility, shows a characteristic maximum slightly above $T_2$ and it also shows a discontinuity in its slope with a sign change (see Figure 11(b)). This behaviour confirms that magnetodielectric effect may be present even in the non-polar antiferromagnetic region $T > T_2$.

The behaviour of the dielectric constant of $Eu_{0.6}Y_{0.4}MnO_3$ under a magnetic field applied along the *a*-axis is associated with a new orientation of the spin configuration that determines a change of the direction of the polarization from *a* to *c*-axis. This means that the AFM2 phase is gradually replaced by the AFM4 phase as the magnetic field increases. As the dielectric constant in the AFM2 phase is much higher than in the AFM4 one, a decrease of the dielectric constant is expected as it is observed in Figure 11(a). Moreover, the small shift of the maximum of the dielectric constant towards higher temperatures under a magnetic field is consistent with the behaviour of the dielectric constant in the AFM4 phase.

The existence of an electric polarization in $Eu_{0.6}Y_{0.4}MnO_3$ samples associated with the magnetic transition at $T_2$, and together with the observed large magnetodielectric effect, implies the existence of an intrinsic magnetoelectric coupling in the polycrystalline samples. Consequently, we have studied in $Eu_{0.6}Y_{0.4}MnO_3$ the magnetic field dependence of electric polarization. Figure 12(a) depicts the time dependence of the electric polarization and the magnetic field. Figure 12(b) shows the magnetic field dependence of the electric polarization, measured at 14 K, obtained from Figure 12(a). As the magnetic field starts to increase, the electric polarization first goes up and reaches its maximum value around 40 kOe, after which it decreases continuously up to 90 kOe. The oscillation of the magnetic field with amplitude of 90 kOe induces oscillation of the electric polarization with maximum amplitude of about 1.5 $\mu C/m^2$ at a double frequency, suggesting the existence of a quadratic magnetoelectric coupling in $Eu_{0.6}Y_{0.4}MnO_3$. The magnetic field dependence of the polarization can be well represented by a butterfly loop shape, as shown in Figure 12(b). In fact, the sign of polarization does not change, but the magnitude of the effect is much smaller than the one reported recently for $Ba_2Mg_2Fe_{12}O_{12}$ by Isiwata *et al.*[19] The decrease of the polarization observed for magnetic fields higher than 40 kOe is due to the onset of the AFM4 phase. The magneto-polarization effects, whose mechanism is related to the modulation of the crystal structure, as it was observed for EuMnO3,[20] $TbMnO_3$ and $DyMnO_3$,[21,22] are likely induced by magneto-elastic interactions, which can be associated with the magnetic field induced transition from the commensurate to incommensurate modulated structure.[8] In order to figure out the validity of such an assertion, we have measured the magnetostriction effect from -90 kOe to 90 kOe, using a modified capacitance dilatometer technique and a PPMS cryostat.[23] To keep the same configuration as that one used for magneto-polarization measurements, the magnetic field was applied perpendicularly to the direction along which the change of dimension was measured. Figure 12(c) shows the strain $\Delta L/L$ as a function of the magnetic field, measured at 14 K. Although a correlation exists between the magnetic field induced polarization changes and the $\Delta L/L$ magnetic loops, the magnetostriction effect is of the order of $10^{-5}$, which is much lower than the order of the magnetodielectric effect, whose value, calculated from our results, is $\sim 2 \times 10^{-1}$. Therefore, in $Eu_{0.6}Y_{0.4}MnO_3$, the effect of a magnetic field on both the dielectric constant and polarization is genuine.

## IV. CONCLUSIONS

From the results presented in this work, we can conclude that $Eu_{0.6}Y_{0.4}MnO_3$ is very easily polarisable under an electric field, and the polarization obtained from integration of the displacement currents, after cooling the sample under an electric field, corresponds to the sum of spontaneous and induced polarizations. This assertion is supported by the study of polarization reversal, which yields a remanent polarization of $\sim 40$ $\mu C/m^2$, which is much lower than the value reported in current literature. The study of hysteresis loops and pyroelectric current obtained by polarization reversal and displacement current measurements, respectively, yields unambiguously that $Eu_{0.6}Y_{0.4}MnO_3$ is ferroelectric between 23 K and 7 K. Above $T_N$, the observed dielectric relaxation, which is well described by an energy level of $0.020 \pm 0.003$ eV, is very probably independent of yttrium content in $Eu_{1-x}Y_xMnO_3$. In fact, a dielectric relaxation study carried out in $EuMnO_3$ ceramics led to a similar value for the activation energy.[15] Contrarily, the energy levels revealed through the study of displacement currents in the temperature range 50 K – 180 K, are surely due to the Yttrium ions. The partial substitution of $Eu^{3+}$ by $Y^{3+}$, and its random distribution probably yield the complex shape of the displacement current in the temperature range 50 K – 180 K. So, the presence of $Y^{3+}$ should be the principal factor that determines the high level of the induced polarization in the $Eu_{1-x}Y_xMnO_3$ system, even at low temperatures.

As it is expected, the displacement current versus temperature is also rather complex, due to a random orientation of the ceramic grains. So, in this system, we can not completely separate the contributions from the electric polarization obtained along *a* and *c*-axes. The study of the displacement currents under different poling electric fields clearly show the importance of induced currents that are different in the AFM2 and AFM4 phases.

We have found clear evidence for the influence of a magnetic field in the polar properties of $Eu_{0.6}Y_{0.4}MnO_3$. The study of electric polarization of $Eu_{0.6}Y_{0.4}MnO_3$ under a magnetic field, at 14 K, yields a value of ~36 $\mu C/m^2$, in fact, of the same order of magnitude of the remanent polarization determined from polarization reversal experiments. By increasing the magnetic field we observe an increase of the polarization, which attains its maximum value by 40 kOe, and then, the polarization decreases by further increasing the magnetic field. Qualitatively, these results are similar to those reported by Noda *et al* [8] and Murakawa *et al*.[9] The latter authors assumed a cycloidal magnetic structure for $Eu_{0.6}Y_{0.4}MnO_3$ crystals, and suggested that the behaviour observed is due to a flexible rotation of the conical spin structure under a magnetic field.

The comparison between the results obtained in polycrystalline samples and single crystals in what concerns the effect of an applied magnetic field on the electric polarization confirms the threshold magnetic field value for the polarization rotation from the *a*- to the *c*-direction, but the shape and magnitude of the response are different. These features can be well understood in the scope of the granular nature of the samples by assuming a random distribution of the direction of crystallographic axes in the sample. Following the results of Noda *et al*,[8] obtained in single crystals (see Figures 1(b) and 1(c) of Ref. 8), we can see that at 14K the magnetic field oriented along the *a*-axis decreases by an amount of 15% the magnitude of the electric polarization along the *a* direction, up to $B_o \sim 35$ kOe, but around this value, there is a sudden decrease of this polarization component, which takes a very low value, while for $B > B_o$, a sudden increase of the *c*-component of the electric polarization takes place, reaching a maximum value, which is about 8 times lower than the *a*-component, for $B < B_o$.[8,9]

In the results displayed in Figure 12(b), we see that the polarization measured in ceramic samples starts to increase with the magnetic field magnitude, attains a maximum value around 40 kOe and then starts to decrease rather slowly. Moreover, the variation of the polarization in the ceramics is much lower than in single crystals.[8,9]

Taking into account the results obtained in single crystals for $B < B_o$, we have not found a plausible explanation for the increase of the electric polarization between 0 Oe and 40 kOe. However, we can understand the slow decrease and the small variation of the electric polarization above 30 kOe. This feature can be well understood if a random distribution of the crystallographic axes in the sample is assumed. If $B_o$ stands for the minimum magnetic field along the *a*-direction that changes the polarization from the *a*-direction to the *c*-direction, and B means the magnitude of the magnetic field in one direction, the change of the *a*-component of the electric polarization ($\Delta P_a$) under the B field is:

$$\Delta P_a = 0 \text{, for B} < B_o,$$

$$\Delta P_a = \frac{P_a}{2}\left[1 - \frac{B_o^2}{B^2}\right] \text{, for B} > B_o. \qquad (3)$$

The above result means that the electric polarization decreases as $1/B^2$, for $B > B_o$, which describes quite well our results, obtained for $B > 40$ kOe. As in our experiment we have $B_o = 40$ kOe and $B_{max} = 90$ kOe, the maximum variation of polarization is about 2 $\mu C/m^2$, which is of the same order of magnitude observed in this work. This result evidences clearly the effect of the granular nature of the samples in the magnetic field dependence of the electric polarization. However, in the granular samples, the magnitude of $B_o$ is clearly similar to the one found in single crystals. It is worth to stress that when comparing our results and those previously obtained for the corresponding single crystals no differences could be ascertained to their dielectric, magnetic, and magnetoelectric critical behaviour. Instead though, differences could be perceived both in the critical temperatures values and maximum amplitude values of the studied physical properties. These differences stem actually from the granular nature of the corresponding ceramics samples.

**AKCNOWLEDGMENTS**

This work was supported by Fundação para a Ciência e Tecnologia, through the Project PTDC/CTM/67575/2006, STREP MACOMUFI and C'NANO Programs, and by Program Alβan, the European Union Program of High Level Scholarship of Latin America (scholarship no. E06D100894BR).

**Captions**

Figure 1. Phase sequence of $Eu_{0.6}Y_{0.4}MnO_3$. [3,5]

Figure 2. Temperature dependence of the real part (a) and imaginary part (b) of the dielectric constant measured at several fixed frequencies. (c) Temperature dependence of the induced specific magnetic moment, measured under an applied magnetic field of 50 Oe. Curve I: heating run after cooling the sample under zero magnetic field; Curve II: heating run after cooling the sample under an applied magnetic field of 50 Oe. The vertical dashed lines signalize the phase transition temperatures.

Figure 3. Natural logaritm of the relaxation time as a function of 1/T.

Figure 4. P(E) relations recorded at 330 mHz, for different fixed temperatures.

Figure 5. Temperature dependence of the remanent polarization, obtained from the P(E) relations.

Figure 6. Displacement current as a function of the temperature, recorded in heating run after cooling the sample under an electric field, applied between 150 K and 7 K. The heating temperature rate was 2 K/min. Inset: detail of the displacement current in the temperatuyre range 10 K – 30 K.

Figure 7. Displacement current density as a function of the temperature, recorded in heating run, with different temperature rates, after cooling the sample under an external electric field, applied between 200 K and 7 K.

Figure 8. Equilibrium polarization and activation energy as a fucntion of temperature, obtaiend from the best fit of Equation (2) to the experimental data above 50 K.

Figure 9. (a) Displacement current density as a function of the temperature, recorded in heating run, after cooling the sample under an external electric field, applied between 40 K and 7 K. (b) Electric polarization as a function of the temperature, obtained from time integration of the displacement current density displayed in Figure 9(a). The inset of Figure

9(b) compares the remanent polarization shown in Figure 4 with the polarization obtained by time integration of the displacement current, measured in heating run after cooling the sample under the lowest applied polarizing electric field (1 kV/m).

Figure 10. (a) Displacement current density as a function of the temperature, recorded in cooling run (temperature rate 2 K/min) between 40 K and 7 K, after cooling the sample under an external electric field of 408 kV/m, applied between 200 K and 40 K, following by a heating run (temperature rate 2 K/min) from 7 K to 200 K. (b) Displacement current density as a function of the temperature, recorded in heating run (temperature rate 2 K/min) between 7 K and 200 K, after cooling the sample under an external electric field of 408 kV/m, applied between 40 K and 7 K.

Figure 11. (a) Temperature dependence of the real part of the complex dielectric constant measured at 100 kHz under different magnetic fields. (b) Magnetic field induced change in the real part of the complex dielectric constant as a function of the temperature.

Figure 12. (a) Electric polarization of $Eu_{0.6}Y_{0.4}MnO_3$, recorded at 14 K(left scale), and magnetic field as a function of time (right scale). (b) Electric polarization as a function of magnetic field, measured at 14 K. (c) Transversal magnetostriction as a function of magnetic field, measured at 14 K.

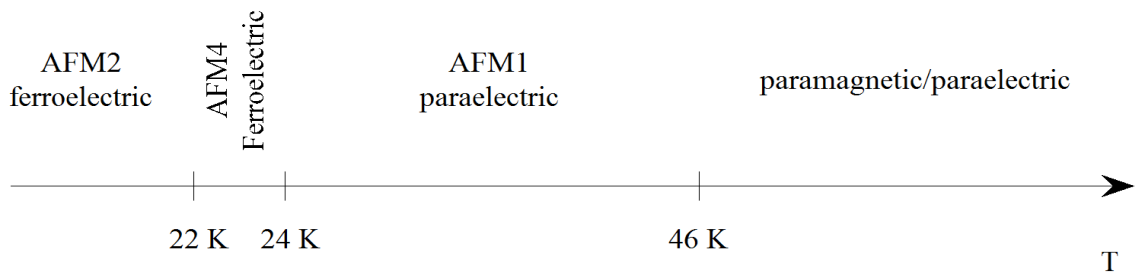

Figure 1

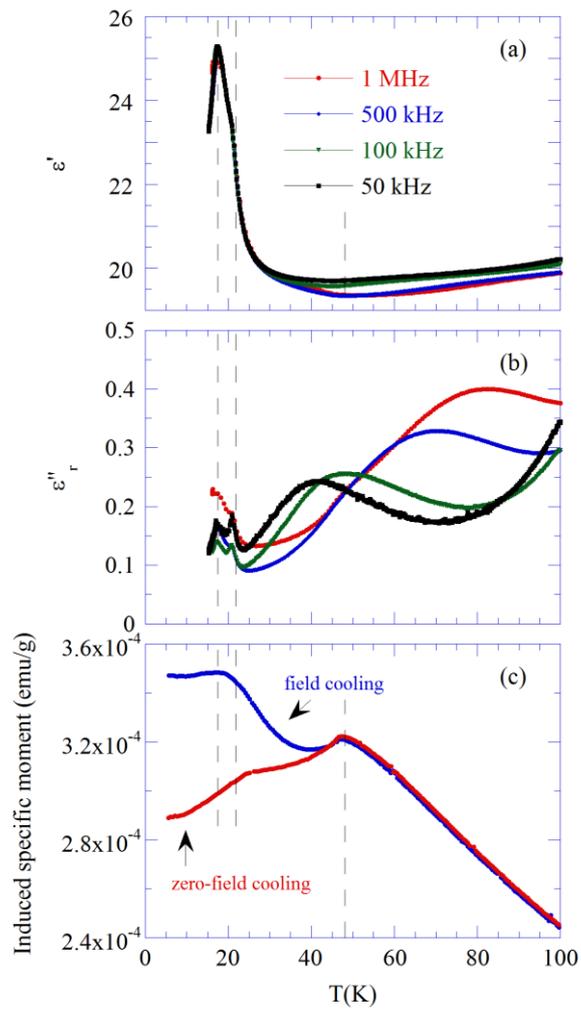

Figure 2

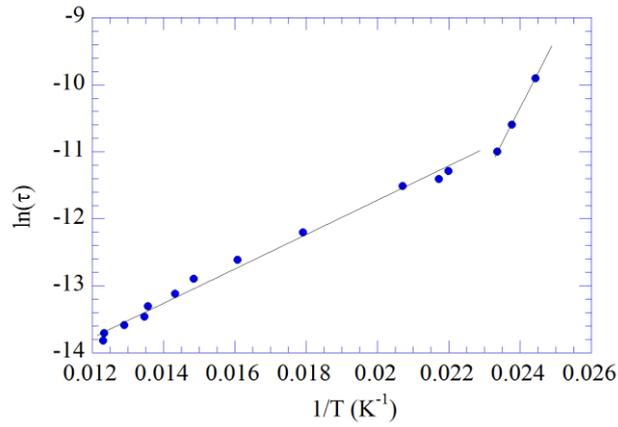

Figure 3

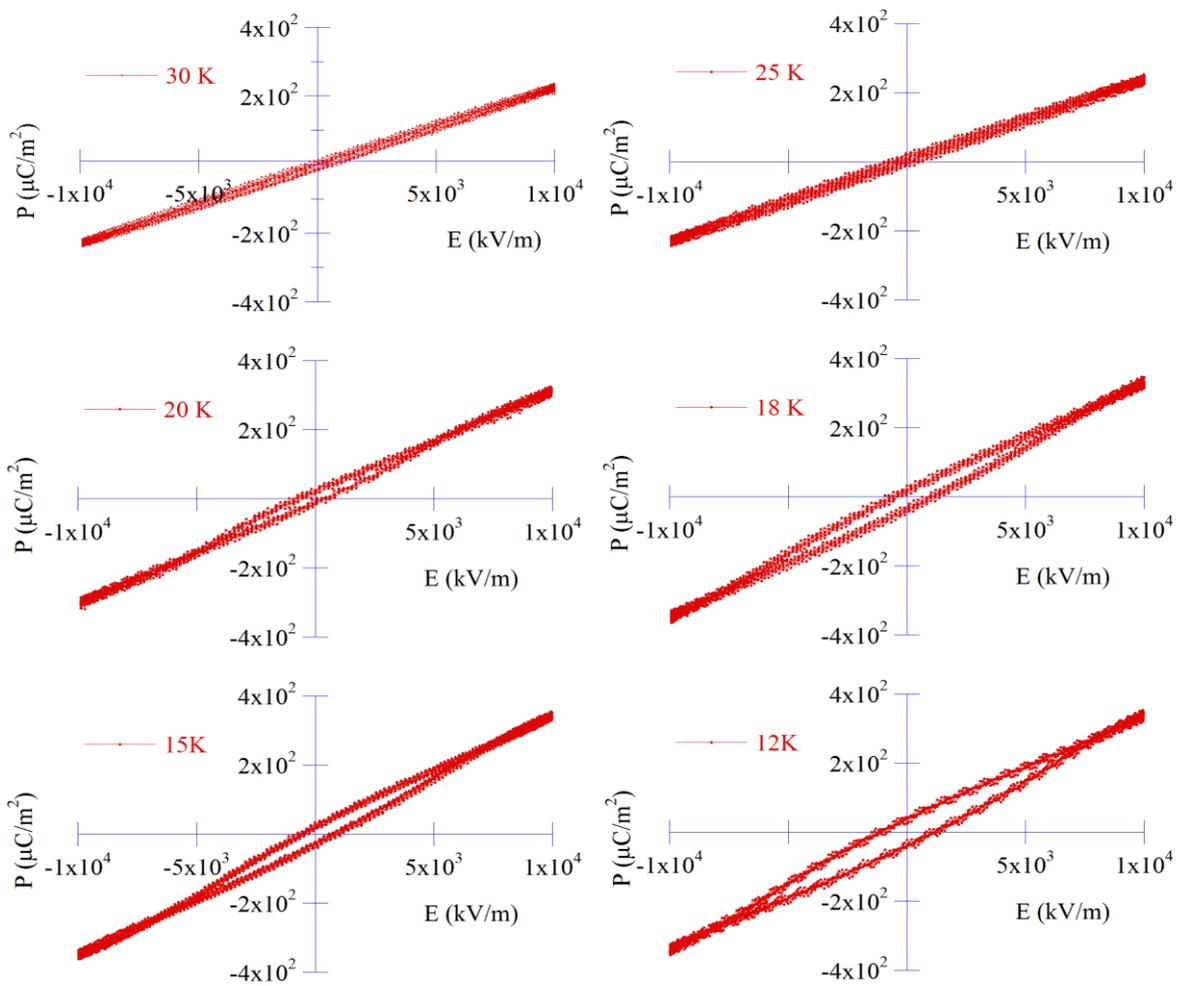

Figure 4

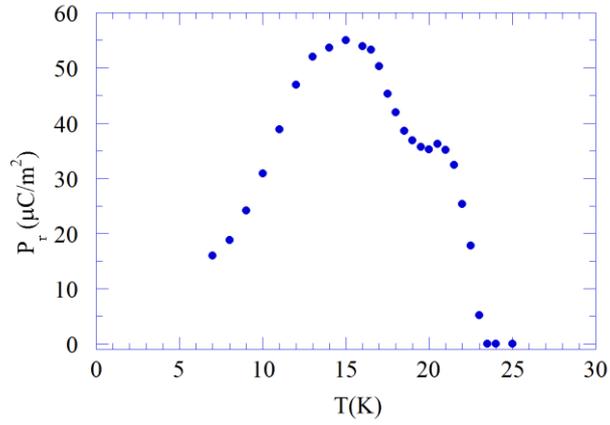

Figure 5

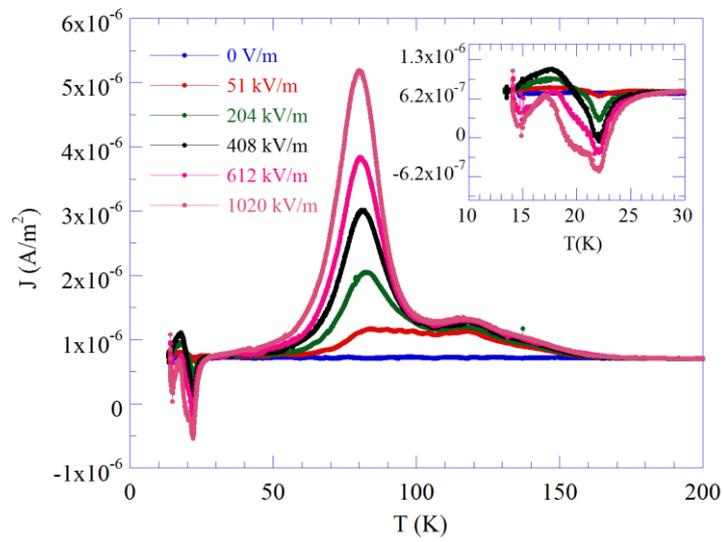

Figure 6

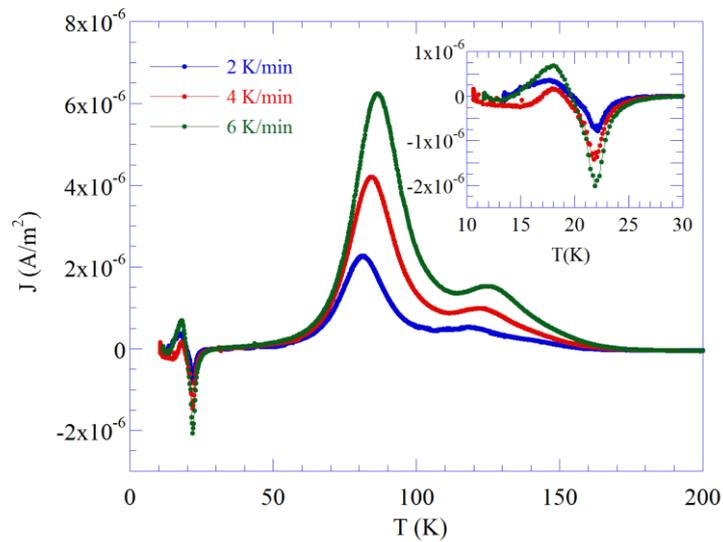

Figure 7

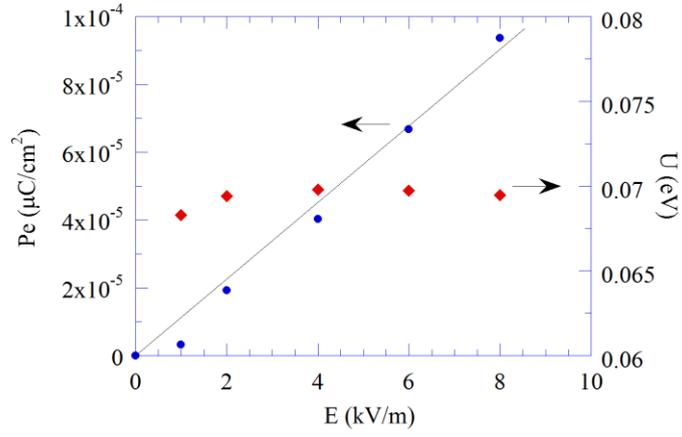

Figure 8

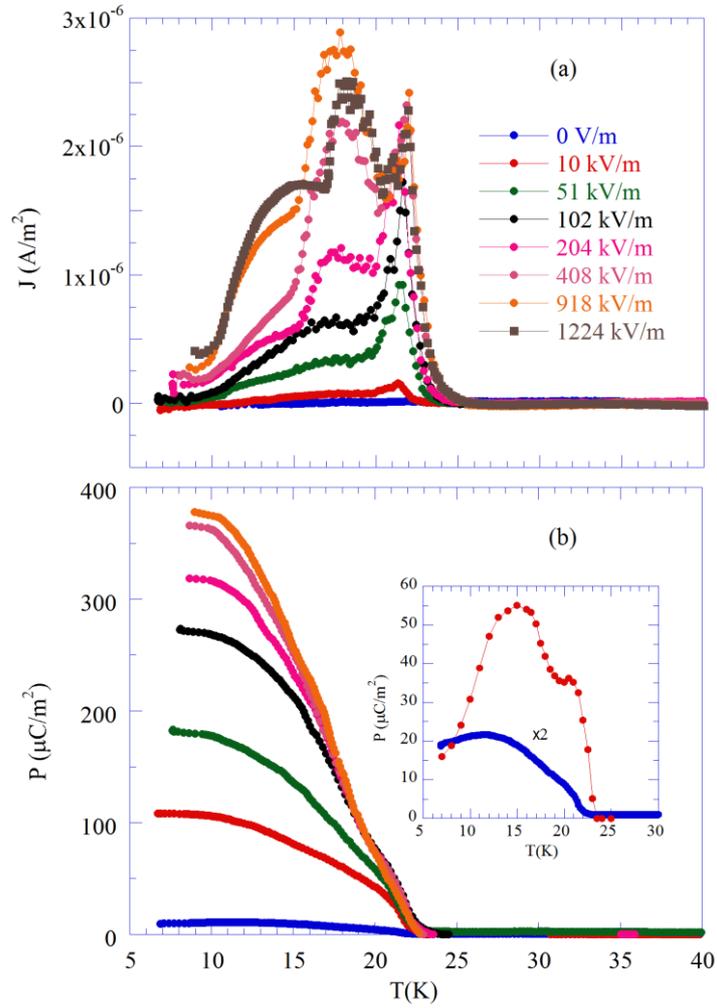

Figure 9

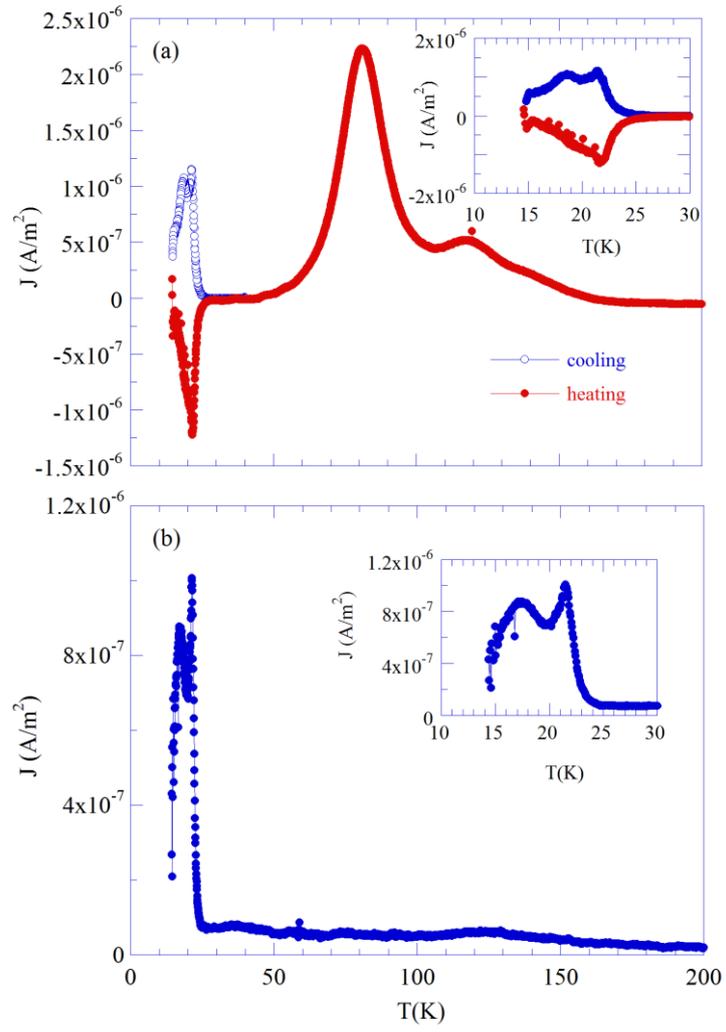

Figure 10

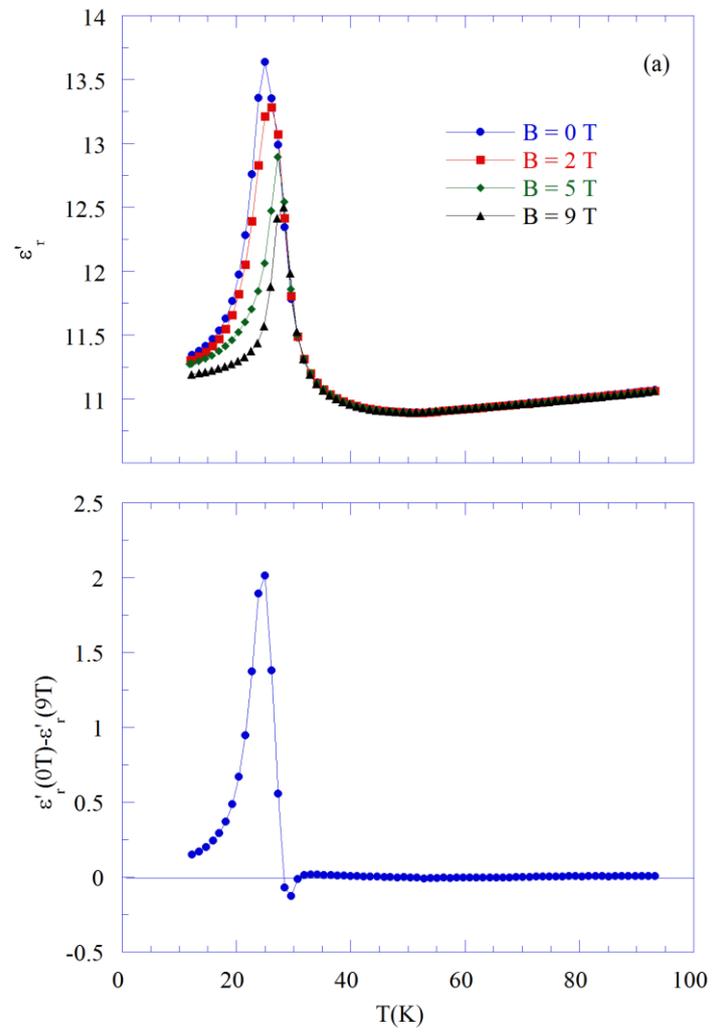

Figure 11

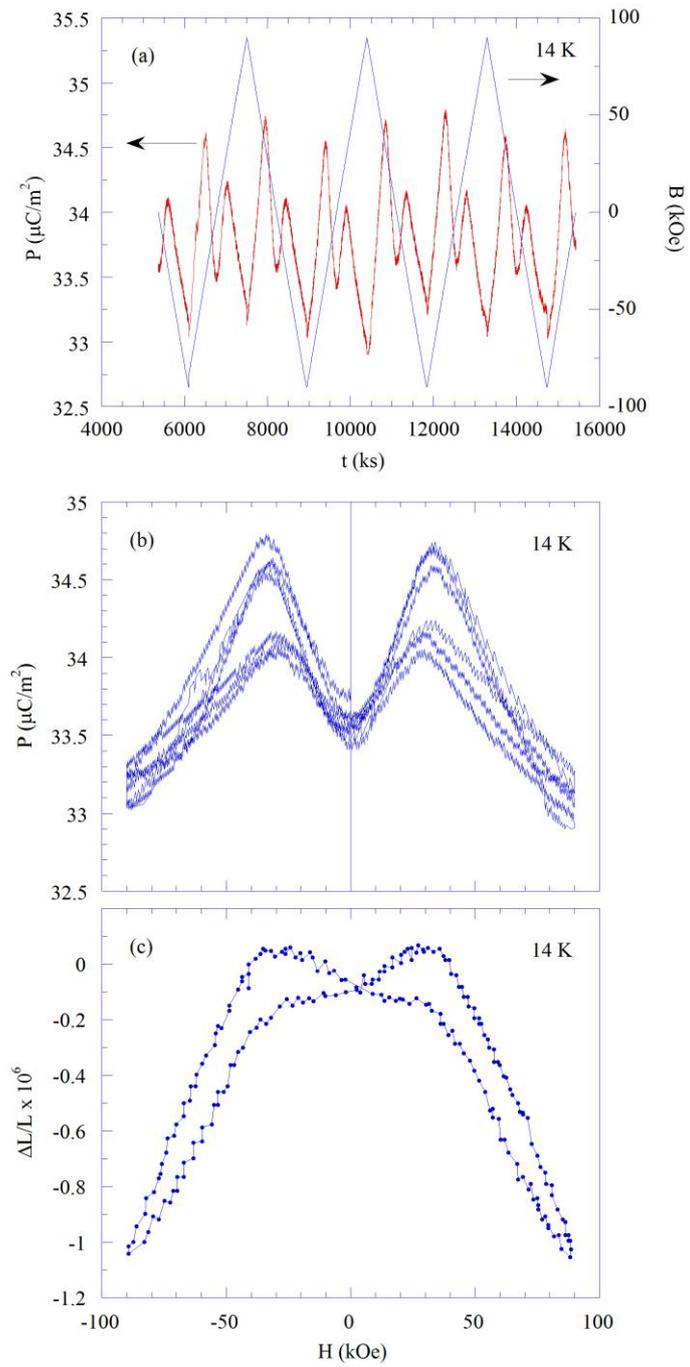

Figure 12